\documentclass[lettersize,journal]{IEEEtran}
\usepackage[T1]{fontenc}
\usepackage[utf8]{inputenc}
\usepackage{amsmath,amsfonts}
\usepackage{algorithmic}
\usepackage{algorithm}
\usepackage{array}
\usepackage[caption=false,font=normalsize,labelfont=sf,textfont=sf]{subfig}
\usepackage{textcomp}
\usepackage{stfloats}
\usepackage{url}
\usepackage{verbatim}
\usepackage{graphicx}
\usepackage{cite}
\usepackage{color}
\usepackage{colortbl}
\usepackage{lmodern} 

\definecolor{lightgray}{gray}{0.9}
\begin{document}
\title{Khmer Semantic Search Engine (KSE): Digital Information Access and Document Retrieval}

\author{Nimol Thuon,~\IEEEmembership{Member,~IEEE,}
\thanks{This work extends the Khmer semantic search engine project developed in 2016 by author N. Thuon. The project was supported by ARES-CCD (Belgium) and the Institute of Technology of Cambodia (ITC). Special thanks go to Dr. Chhun Sophea for supervision and to the ITC students for their data contributions.}
}

\markboth{Khmer Semantic Search Engine: Digital Information Access and Document Retrieval}%
{Shell \MakeLowercase{\textit{et al.}}: Bare Demo of IEEEtran.cls for IEEE Transactions on Magnetics Journals}

\maketitle

\begin{abstract}

The search engine process is crucial for document content retrieval. For Khmer documents, an effective tool is needed to extract essential keywords and facilitate accurate searches. Despite the daily generation of significant Khmer content, Cambodians struggle to find necessary documents due to the lack of an effective semantic searching tool. Even Google does not deliver high accuracy for Khmer content. Semantic search engines improve search results by employing advanced algorithms to understand various content types. With the rise in Khmer digital content—such as reports, articles, and social media feedback—enhanced search capabilities are essential. This research proposes the first Khmer Semantic Search Engine (KSE), designed to enhance traditional Khmer search methods. Utilizing semantic matching techniques and formally annotated semantic content, our tool extracts meaningful keywords from user queries, performs precise matching, and provides the best matching offline documents and online URLs. We propose three semantic search frameworks: semantic search based on a keyword dictionary, semantic search based on ontology, and semantic search based on ranking. Additionally, we developed tools for data preparation, including document addition and manual keyword extraction. To evaluate performance, we created a ground truth dataset and addressed issues related to searching and semantic search. Our findings demonstrate that understanding search term semantics can lead to significantly more accurate results.
\end{abstract}


\section{Introduction}
\IEEEPARstart{T}{he} rapid growth of digital content has significantly increased the need for efficient information retrieval systems. Search engines play a critical role in this process by helping users find relevant information across various fields such as education, health, and entertainment \cite{ref1}. However, traditional search engines often rely on syntactic matching, which fails to capture the nuanced meanings of different terms. This limitation is particularly evident in the context of Khmer language content, where users frequently struggle to find the documents they need due to the lack of effective semantic search tools \cite{ref2}.

\begin{figure}[!t]
\centering
\includegraphics[width= 3.5 in]{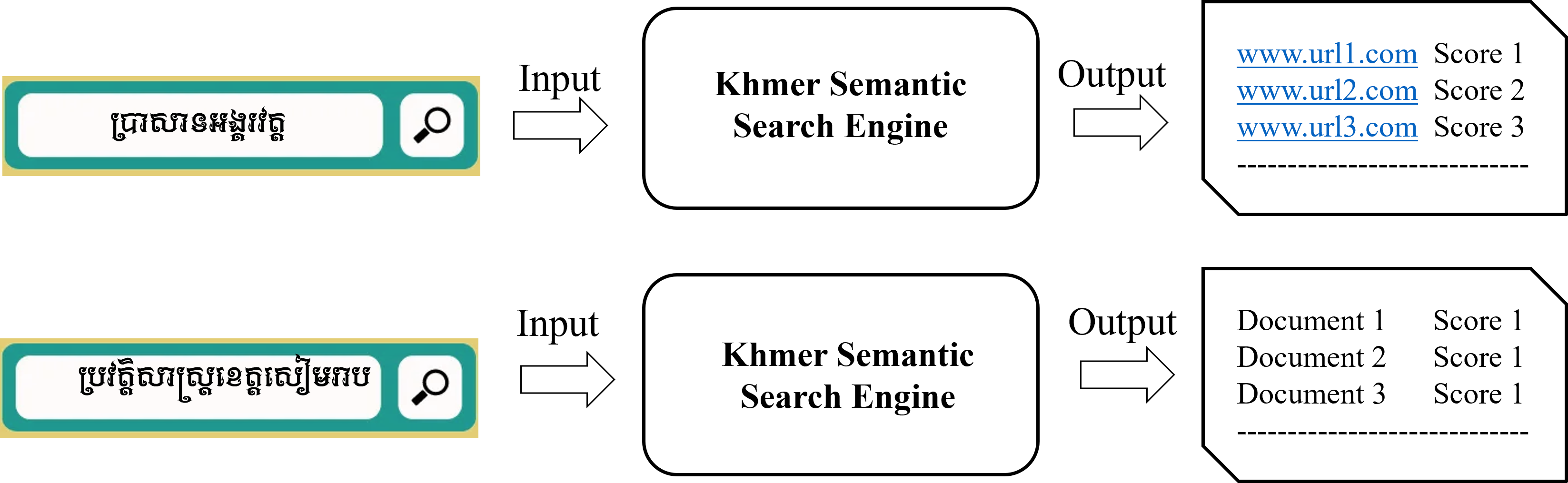}
\caption{Khmer semantic search engine in global view.}
\end{figure}

In Cambodia, the generation of Khmer digital content such as reports, articles, and social media feedback is on the rise. Despite this, the absence of a robust Khmer semantic search engine means that users often cannot access the most relevant documents. Even major search engines like Google do not yet provide high accuracy for Khmer language searches, highlighting a significant gap in the available technology \cite{ref3}. Khmer documents, both historical and modern, exhibit complex writing forms and grammar, presenting additional challenges [12][15].

The objective of this research is to address this gap by developing the Khmer Semantic Search Engine (KSE). This tool aims to enhance traditional Khmer search methods by utilizing advanced semantic matching techniques. By formally annotating semantic content and extracting meaningful keywords from user queries, KSE can perform precise matching and deliver the best matching URLs.

This paper presents the research design, implementation, and evaluation of the Khmer Semantic Search Engine (KSE). It proposes three semantic search frameworks: semantic search based on a keyword dictionary, semantic search based on ontology, and semantic search based on ranking.

Additionally, it describes the development of tools for data preparation, including document addition and manual keyword extraction. The evaluation of KSE's performance, using a ground truth dataset, demonstrates how understanding the semantics of search terms can significantly improve search accuracy.

This research contributes to the field in several significant ways:

\textbf{1. Development of a Khmer Semantic Search Engine:} We introduce a new search engine specifically tailored for the Khmer language, addressing the unique challenges associated with Khmer content retrieval.

\textbf{2. Advanced Semantic Matching Techniques:} The research incorporates sophisticated semantic matching techniques that go beyond simple keyword matching, enhancing the accuracy of search results.

\textbf{3. Comprehensive Data Preparation Tools:} We developed tools for data preparation, including automatic and manual keyword extraction, and ontology-based semantic enrichment to ensure high-quality input data for the search engine.

\textbf{4. Evaluation with Ground Truth Dataset:} The performance of KSE is rigorously evaluated using a ground truth dataset, providing a reliable benchmark for assessing its effectiveness.

In the following sections, we will delve into the introduction, objectives, and challenges associated with building an effective Khmer semantic search engine. We will also review related literature, propose the KSE architecture, and discuss the implementation and evaluation methodologies. Finally, we will conclude with the findings and future perspectives of this research.

\section{RELATED WORKS}

Semantic search has been a focal area of research, particularly for languages with a rich vocabulary and complex syntax. Various approaches have been explored to enhance the accuracy and relevance of search results by understanding the semantic context of queries and documents.
\subsection{Ontology-Based Approaches}

\begin{figure}[!t]
\centering
\includegraphics[width= 3.3 in]{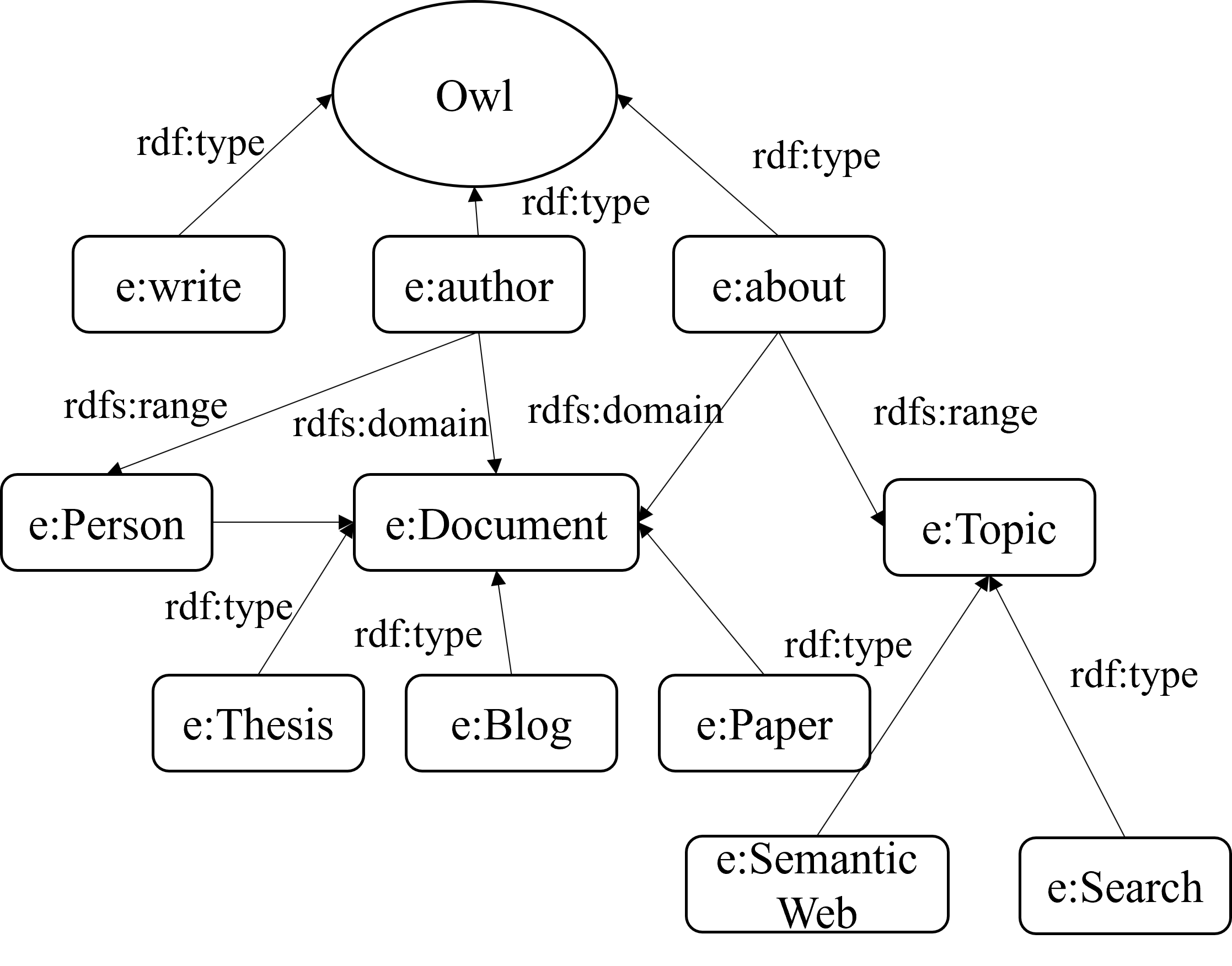}
\caption{A simple search engine ontology.}
\end{figure}

Ontologies play a crucial role in representing domain knowledge and facilitating semantic matching. Gruber defines an ontology as “a specification of a representational vocabulary for a shared domain of discourse—definitions of classes, relations, functions, and other objects” \cite{ref4}. Ontologies have been extensively used to improve search engines by enabling more precise query interpretations and document classifications. Chandrasekaran et al. also emphasize the importance of ontologies in defining the properties and relationships of objects within a specific domain \cite{ref5}.
\begin{figure}[!t]
\centering
\includegraphics[width= 3.3 in]{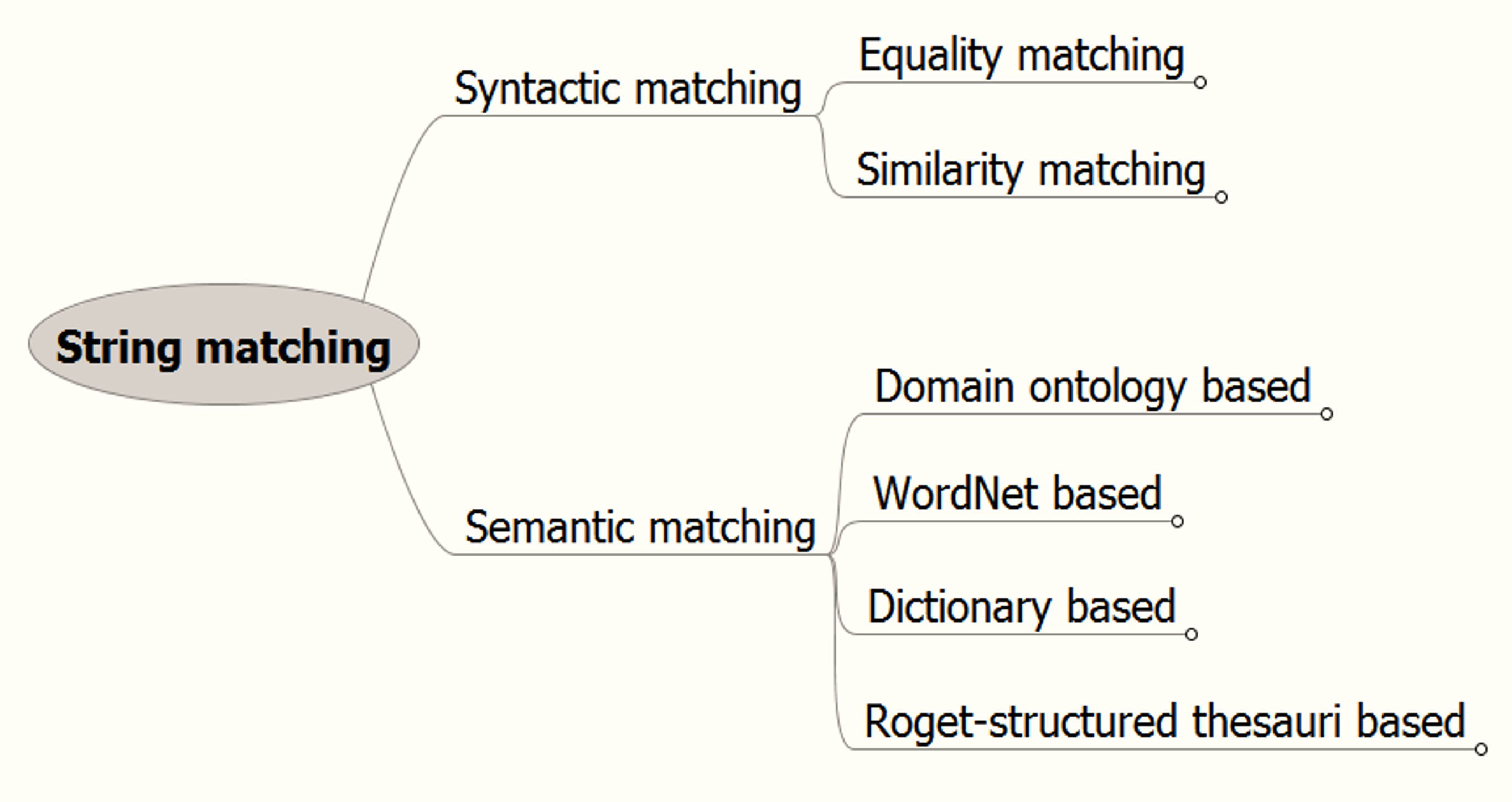}
\caption{String matching techniques}
\end{figure}

An ontology consists of three elements: vocabularies, explicit specifications, and constraints. Vocabularies describe an ontology domain, and constraints capture knowledge about a domain. Objects in the ontological concept are generally represented by nouns, and relationships are usually specified by verbs in a sentence.

For example in Fig. 2, a simple ontology of a search engine can include vocabularies such as search, author, write, person, document, thesis, paper, topic, semantic web, and search. A search engine is specified by properties such as query, results, relevance score, and search time. Relationships can be defined as well, such as Author, Document, Thesis, Paper, Topic, Semantic Web, and Search.

The cardinality defines that a search engine processes multiple queries, each returning multiple results. This can be expressed in ontological terms as "a search engine performs Search a query" and "a query returns multiple results." (see detail Fig. 3)

\subsection{Keyword Extraction Methods}

Keywords are essential for indexing and efficiently retrieving documents. Several methods have been proposed for automatic keyword extraction, ranging from simple statistical approaches to more complex machine learning techniques. One such statistical method is Term Frequency-Inverse Document Frequency (TF-IDF), which evaluates the importance of a term based on its frequency in a document relative to its frequency across all documents \cite{ref6}.

Frank et al. introduced the Keyphrase Extraction Algorithm (KEA), a widely used tool that combines statistical and machine learning methods for automatic keyword extraction \cite{ref6}. This approach has been effective in identifying key terms that represent the main content of documents. Additionally, Hulth demonstrated that incorporating linguistic knowledge significantly improves the accuracy of keyword extraction by considering the syntactic and semantic structure of the text.

In the context of Khmer language, previous studies have proposed methods for extracting main keywords from Khmer text using multi-step processes involving a stop word corpus dictionary \cite{ref8}. These approaches are designed to handle the unique challenges posed by the Khmer language, including its complex script and lack of spaces between words, making the extraction process more accurate and relevant.

\subsection{Semantic Matching}

Various techniques have been developed to perform semantic matching between queries and documents. These techniques aim to understand the meanings and relationships of words to improve the accuracy of search results.

\textbf{WordNet}: A widely used lexical database of English, WordNet groups words into sets of synonyms called synsets and provides short definitions and usage examples. It also records various semantic relationships between these synonym sets. This resource has been instrumental in measuring semantic similarity between words \cite{ref10}.

\textbf{Wu and Palmer Algorithm}Wu and Palmer Algorithm: Wu and Palmer proposed an algorithm that measures semantic similarity by considering the depths of synsets in the WordNet taxonomy and their least common subsumer (LCS) \cite{ref11}. The similarity score is calculated based on how closely related the synsets are within the hierarchical structure of WordNet \cite{ref10}. This method is particularly effective in capturing the hierarchical nature of semantic relationships.

\textbf{Information-Theoretic Approaches}: Resnik introduced an information-theoretic approach that uses corpus statistics to determine the similarity of two words \cite{ref13}. This method involves calculating the information content of the words and their shared ancestors in a taxonomy, providing a probabilistic measure of similarity. By leveraging large text corpora, this technique can effectively quantify the degree of relatedness between words based on their contextual usage \cite{ref14}.

\textbf{Word Embeddings}: Modern approaches often involve word embeddings, such as Word2Vec, GloVe, and BERT, which represent words in continuous vector spaces. These embeddings capture semantic relationships based on word co-occurrence patterns in large corpora, enabling more sophisticated matching techniques \cite{ref16}. These models can understand context, disambiguate word meanings, and recognize synonyms, thus improving the relevance of search results.

\textbf{Latent Semantic Analysis (LSA)}: Another important technique is Latent Semantic Analysis, which reduces the dimensionality of the term-document matrix using singular value decomposition (SVD). This method identifies patterns in the relationships between terms and documents, allowing for the capture of underlying semantic structures \cite{ref13}.

\textbf{Ontology-Based Matching}: Ontologies are also used for semantic matching by providing a structured representation of domain knowledge. They define classes, relations, and properties of concepts within a specific domain, allowing for more precise query interpretations and document classifications \cite{ref17}. Ontology-based approaches can integrate domain-specific knowledge into the matching process, enhancing the accuracy and relevance of search results \cite{ref18}.

\section{CHALLENGES IN KHMER LANGUAGE PROCESSING}
The development of semantic search tools for the Khmer language faces several significant challenges. While substantial progress has been made in semantic search technologies for widely spoken languages like English, these advancements have not yet been fully realized for Khmer. This disparity can be attributed to several factors:

\textbf{Complexity of Khmer Syntax}: The Khmer language features a complex syntactic structure that includes a wide array of compound words, 
intricate grammatical rules, and context-dependent meanings. This complexity makes it difficult to develop accurate natural language  processing (NLP) tools that can effectively parse and understand Khmer text \cite{ref19}.

\textbf{Limited Research and Development}: There has been relatively little research focused on Khmer language processing compared to more widely spoken languages. This lack of attention results in fewer advancements and slower progress in developing effective NLP tools for Khmer. Existing systems often suffer from low accuracy and fail to meet the needs of Khmer-speaking users \cite{ref20}.

\textbf{Diverse Dialects and Regional Variations}: Khmer is spoken in various dialects across different regions of Cambodia. These dialectal differences add another layer of complexity to the development of a universal semantic search tool that can understand and process all variations accurately \cite{ref21}.

\textbf{Inadequate Support for Khmer in Major Search Engines}: Major search engines like Google have not yet achieved high accuracy for Khmer content. The lack of support from these platforms means that Khmer-speaking users do not have access to efficient search tools, further emphasizing the need for specialized solutions \cite{ref22}.

\section{KHMER SEMANTIC SEARCH ENGINE}
\begin{figure*}[!t]
\centering
\includegraphics[width = 6.7 in]{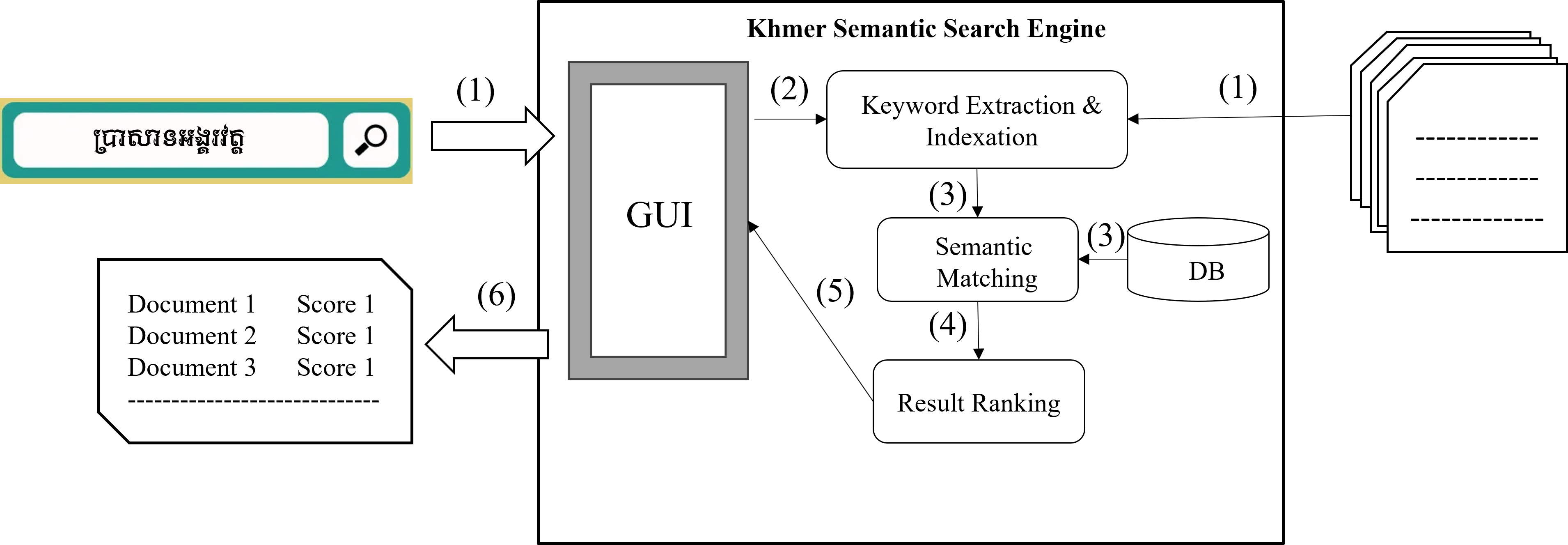}
\caption{Khmer semantic search engine overview (with document indexation)}
\end{figure*}
The Khmer Semantic Search Engine (KSE) is designed to address the unique challenges of processing and retrieving Khmer language content. As shown in Fig. 4, the architecture of KSE includes several key components and processes that work together to provide accurate and relevant search results process and as following:
\subsection{Models of Offline and Online Processing}
\textbf{1) Offline Processing (Figure 4)}: In the offline processing model, the system performs matching with offline documents stored in a database. This approach is particularly useful for stable and unchanging content, such as historical documents and official reports. Users or administrators begin by uploading documents into the system. Once uploaded, the system extracts significant keywords from these documents and indexes them in the database. This indexed information allows for efficient retrieval when users perform searches. When a user enters a search query via the graphical user interface (GUI), the system matches the query keywords with the indexed document keywords using ontology-based matching and other techniques. The results are then ranked based on relevance scores, which consider the semantic similarity between the query and the document keywords. Finally, the ranked search results are displayed to the user, with the most relevant documents appearing first.

\textbf{2) Online Web Page Processing 
(Figure 5)}: In the online web page processing model, the system performs real-time matching of user queries with online web pages, allowing for up-to-date information retrieval such as the latest news articles and social media posts. The system continuously crawls the web to gather new content from various sources, including news sites, blogs, and social media platforms. As new content is discovered, the system extracts keywords and indexes these web pages in real-time. When a user enters a search query via the GUI, the system matches the query keywords with both the indexed offline documents and the newly indexed  online content. Advanced semantic matching techniques are used to ensure accurate matches. The results are then ranked based on their relevance, taking into account both offline documents and online web pages. The ranked results are displayed to the user, showcasing both types of content with the most relevant results at the top.

\subsection{Detailed Processes}
\begin{figure*}[!t]
\centering
\includegraphics[width = 6.7 in]{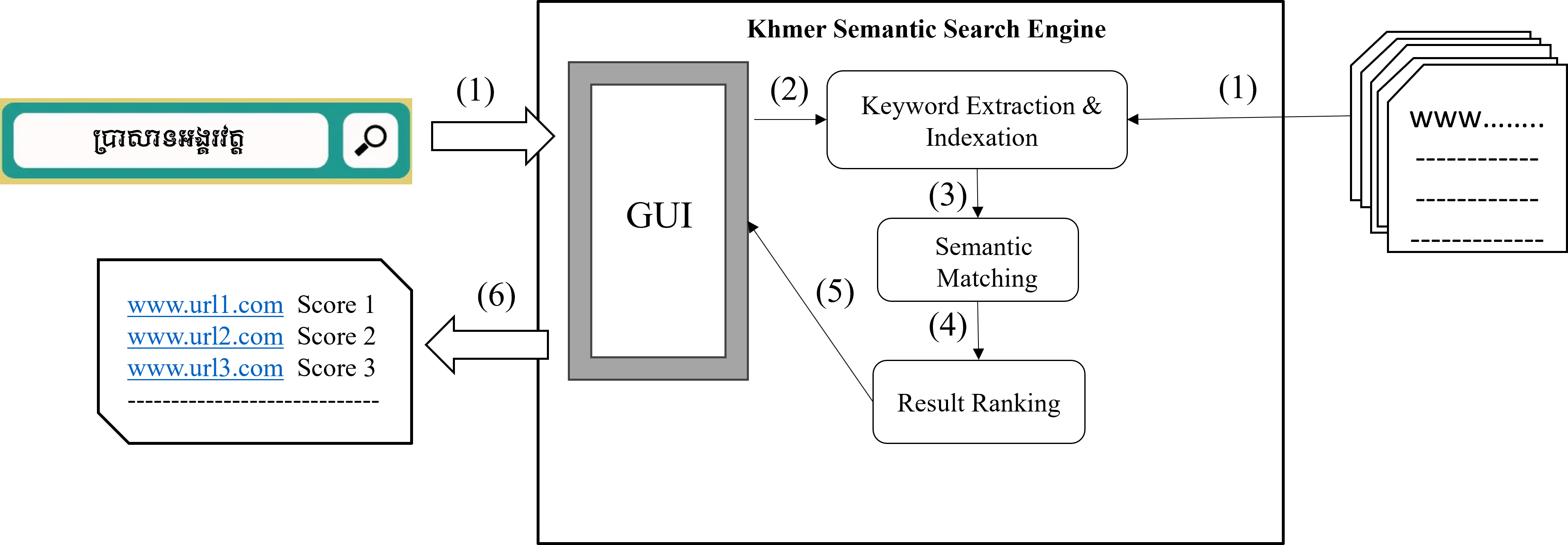}
\caption{Khmer semantic search engine overview (with web page processing)}
\end{figure*}
The semantic search engine can process both indexed documents in the database (see Figure 4) and perform real-time searches for available documents online (see Figure 5). The processes involved are detailed below:\\
\begin{figure}[!t]
\centering
\includegraphics[width= 3.5 in]{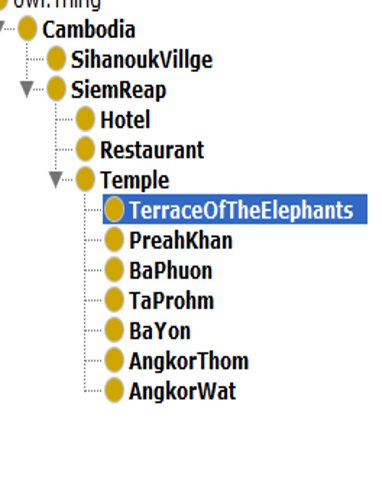}
\caption{Ontology-based Tourism Knowledge Graph for Cambodia}
\end{figure}

\textbf{1) Input Modules}

\textbf{Query Input}: Users enter their search queries in Khmer through a user-friendly graphical user interface (GUI). This input is crucial for initiating the search process. The GUI is designed to be intuitive and accessible, encouraging users to engage with the search engine without any technical difficulties. 
For example, a user might enter the query "Cambodia".

\textbf{Document Input}: Documents, including reports, articles, and social media feedback, are added to the system. These documents are processed and indexed for efficient retrieval. The input can come from various sources, such as uploading files, entering text directly, or scraping web content. This ensures comprehensive coverage of content.\\

\textbf{2) Output Modules}

\textbf{Search Results}: The system outputs the most relevant search results based on the user's query. These results include URLs and a relevance score, providing users with a clear indication of the usefulness of each result. For instance, if the query is "Cambodia" the results might include links to government reports, historical articles, and relevant social media posts.

\textbf{Result Ranking}: The search results are ranked based on their relevance to the query. The relevance scores are calculated by considering the semantic similarity between the query and document keywords. The ranked search results are then displayed to the user through the GUI. The most relevant results appear at the top, helping users quickly find the information.

\subsection{Semantic Search Matching Process}

The semantic search matching process in our Khmer Semantic Search Engine (KSE) involves three main components: semantic search based on keywords, semantic search based on tourism ontology for Cambodia, and semantic search based on ranking. Each component is designed to enhance the accuracy and relevance of the search results through detailed and technically robust processes.\\

\textbf{1)Semantic Search Based on Keywords}\\

The first component involves a sophisticated keyword extraction process from user queries. This begins with preprocessing steps such as:

\textbf{Tokenization:} The query is split into individual words or tokens. For example, the query "best cultural sites in Phnom Penh" is tokenized into ["best", "cultural", "sites", "in", "Phnom", "Penh"].

\textbf{Stop-Word Removal:} Common Khmer stop words are removed to focus on significant terms. This step uses a predefined list of stop words specific to the Khmer language, as detailed in \cite{ref8}. For instance, Khmer stop words like (in), (of), (has), and (is) are filtered out. Example: the query (waterfalls in Phnom Penh) would be reduced to [(waterfalls), (Phnom Penh)].

\textbf{Stemming and Lemmatization:} Words are reduced to their base or root form. For instance, "sites" becomes "site". This step ensures that variations of a word are considered equivalent.

\textbf{TF-IDF Calculation:} Term Frequency-Inverse Document Frequency (TF-IDF) is used to weigh the importance of each term within the query relative to its frequency in the document corpus. 

\textbf{Keyword Matching:} The extracted and weighted keywords are then used to search the indexed documents. Each document in the database has been pre-processed in a similar manner, with keywords extracted and stored in an index for fast retrieval. The system compares the keywords from the user query with those in the document index to find potential matches.\\

\textbf{2) Semantic Search Based on Tourism Ontology for Cambodia}\\

The second component leverages a domain-specific ontology to enhance the semantic understanding of user queries. The tourism ontology for Cambodia includes detailed information about various tourist destinations, cultural landmarks, historical sites, and other points of interest in Cambodia. The ontology provides a structured representation of domain knowledge, including the relationships between different entities.

The ontology is developed using the Web Ontology Language (OWL) and consists of multiple classes and properties that capture the complexities of Cambodian tourism. Specifically, the ontology includes:

\textbf{Entities:} These include tourist attractions (e.g., Angkor Wat, Royal Palace), types of attractions (e.g., temples, museums), and locations (e.g., Phnom Penh, Siem Reap). Each entity is described with properties such as historical significance, visitor reviews, operational hours, and geographical coordinates.

\textbf{Relationships:} Relationships define how entities interact. For example, "Angkor Wat" (entity) is a "temple" (type) located in "Siem Reap" (location). The ontology also includes hierarchical relationships, such as "Siem Reap" is a part of "Cambodia".

\textbf{Ontology-Based Query Expansion:} When a user enters a query, the system uses the ontology to expand the query semantically. For instance, a query for "temples in Phnom Penh" might expand to include synonyms and related terms like "wats" and specific temple names in Phnom Penh. The system leverages synonyms and related entities defined in the ontology to ensure comprehensive query coverage.

\textbf{Ontology Details:} The ontology comprises over 500 entities, covering major and minor tourist attractions, cultural sites, and events. It includes approximately 1000 relationships that map these entities in various contexts, providing a rich semantic network for query expansion. For example, a query for "temples in Phnom Penh" might be expanded to include "Wat Phnom" and "Wat Botum", both specific temple names in Phnom Penh.

\textbf{Semantic Similarity Calculation:} The system calculates the semantic similarity between the expanded query and the documents using ontology-based metrics. This might involve techniques such as cosine similarity on the vector representations of entities and their relationships, ensuring that documents related to the context of the query are prioritized.\\

\textbf{3) Semantic Search Based on Ranking}\\

The final component focuses on ranking the retrieved documents to present the most relevant results to the user. This process combines both keyword-based and ontology-based relevance and involves several steps:

\textbf{Relevance Scoring:} Documents are scored based on both keyword relevance and semantic relevance. Keyword relevance is determined by the presence and frequency of query keywords in the document, as calculated using TF-IDF scores. Semantic relevance is assessed by comparing the ontology-based context of the query with the document content. This involves measuring the semantic similarity between the entities and relationships identified in the query and those in the documents.

\textbf{Weighted Scoring Algorithm:} The overall score for each document is a weighted combination of keyword relevance and semantic relevance. This approach ensures that both the direct relevance of keywords and the deeper semantic connections are considered, providing a balanced and accurate ranking of documents.

\textbf{Popularity Metrics:} In addition to relevance scores, documents are also ranked based on popularity metrics such as click-through rates, user ratings, and the number of views. These metrics provide insight into how users interact with the documents, indicating their perceived value and relevance. To incorporate these metrics into the final ranking score, we normalize the data to ensure comparability. For example, click-through rates might be scaled to a range between 0 and 1, user ratings could be averaged, and view counts normalized based on the highest and lowest values in the dataset. By including popularity metrics, we ensure that frequently accessed and highly rated documents receive an appropriate boost in their ranking, reflecting their utility and relevance to users.

\textbf{Ranking Adjustments:} The system applies final adjustments to the ranking based on additional factors such as the recency of the document, user-specific preferences, and contextual relevance. For instance, more recent documents or those with higher user engagement might receive a slight boost in their ranking. This ensures that the most relevant and timely information is presented to the user.

\textbf{Presentation of Results:} The top-ranked documents are then presented to the user in a list, with the highest-scoring documents appearing first. Each result includes a brief snippet highlighting the relevant keywords and context to help the user quickly identify the most pertinent information. This presentation aims to enhance the user experience by making it easier to find the desired information quickly.

By integrating these three components, the KSE provides a comprehensive and accurate search experience tailored to the specific needs of Khmer language users and the domain of Cambodian tourism. This multi-layered approach ensures that the search engine can handle complex queries and deliver relevant, high-quality results effectively.

\section{EXPERIMENTAL SETUPS}

\subsection{Stop word dictionary}

For implementing Khmer semantic search, we need to first provide a keyword or search query. This is similar to traditional search engines like Google, where the user inputs a keyword or phrase, and the search results are returned based on relevance to that query. 

In the case of Khmer semantic search, the process involves extracting the main keywords or concepts from the article or content being searched. To do this, we can leverage a Khmer stop word dictionary \cite{ref9}. This dictionary contains around 1,000 common Khmer words that are typically considered "stop words" - words that do not carry significant meaning and can be removed from the text without losing the core information.

By removing these stop words from the article or document, we can extract the more meaningful keywords and concepts. These extracted keywords then serve as the basis for the semantic search, allowing the system to understand the core meaning and context of the content and return more relevant and precise search results.
\begin{figure}[!t]
\centering
\includegraphics[width= 3.5 in]{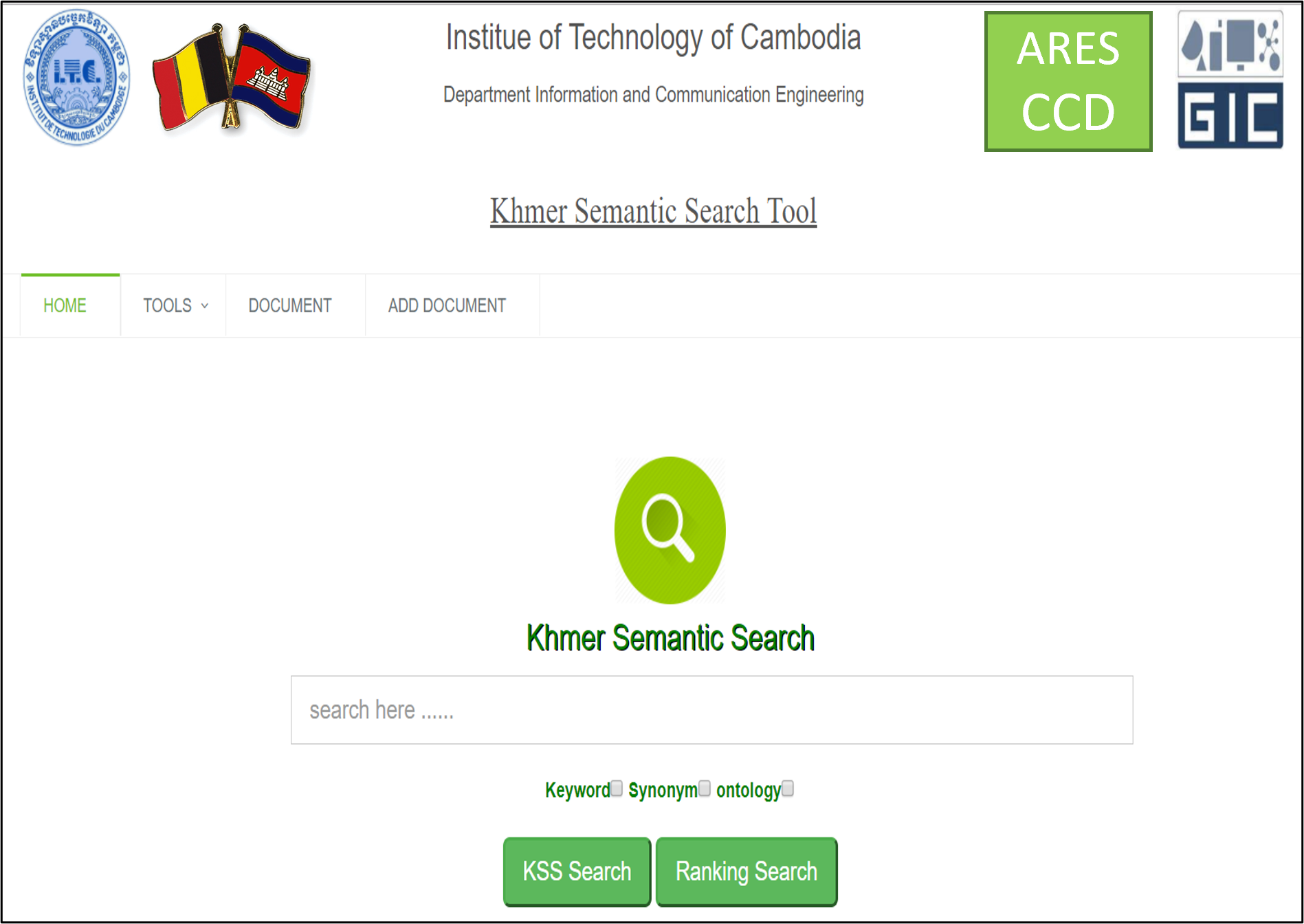}
\caption{Khmer semantic search engine web-based application }
\end{figure}

The use of a Khmer stop word dictionary is a crucial step in the Khmer semantic search process, as it helps to refine the search query and focus on the most important and informative parts of the text. This approach can lead to more accurate and relevant search results, providing users with a more meaningful and efficient search experience when working with Khmer language content.

\subsection{Document contents}

In this case our data is finding the document or article that have the title and content and more over we focus only tourisms. We need the article for making the keyword by using extract keyword.

Because of our data is about getting article that have title and content, so we decided to build the web application by using auto extract title and content of the article that’s need the data from the other website for generate. In this web application, it is simple to use we just input the URL of the article or input link of the website it will generate. After we put the URL then we click generate it will divide two parts are title and description. This is the simple form for generate the article from the website that has place for putting the URL of the article then generate it will show two parts are title and content as in the picture below:

Khmer Semantic search needs a lot of documents for extracting, so we need to ask for help from friends and juniors. So, we developed a web application for hosting that makes it easy to work with. They just put the URL and submit it in this web application. 
Therefore, we get a lot of articles related to the tourism, news, blogs, and social media captions.
Currently, we have 1,150 articles that get from the web application that they submitted.

\subsection{Khmer Text Segmentation}

Khmer segmentation is a crucial step in processing Khmer language documents, as it involves dividing text into meaningful units for further analysis. This section describes both automatic and manual segmentation methods used in the Khmer Semantic Search Engine (KSE).

\textbf{Auto Khmer Segmentation:} Automatic Khmer segmentation \cite{ref24} is employed to facilitate the processing of large volumes of text efficiently. The system utilizes a segmentation algorithm designed specifically for the Khmer language. This algorithm introduces methods to accurately divide text into words and phrases based on linguistic patterns and statistical models. Although automatic segmentation significantly speeds up the processing, it may not always achieve perfect accuracy due to the inherent complexity and contextual nuances of the Khmer language.

\textbf{Manual Segmentation:} Due to the occasional inaccuracies in automatic segmentation, manual segmentation is also employed to ensure high precision, particularly for important documents. In this process, human annotators manually segment the text and extract keywords, ensuring that the segmentation aligns perfectly with the context and meaning of the content.
Manual keyword extraction involves identifying and extracting significant words from both the title and body of the article. This method relies on human expertise to discern the most relevant terms that best represent the document's content.

\subsection{Building ground truth}
To evaluate our methods effectively, we needed a ground truth dataset. This involved manual keyword extraction and identifying the top five ranking documents for various search queries.

We engaged students from the Institute of Technology of Cambodia (ITC) to assist in the manual extraction of keywords and the identification of relevant search results. The process was as follows:

\begin{itemize}
    \item \textbf{Document Selection:} A total of 1,150 documents were selected for testing. These included a mix of Khmer web pages and documents to ensure comprehensive coverage.
    \item \textbf{Keyword Extraction:} Students manually extracted keywords from each document. On average, three keywords were extracted per document, resulting in a total of 9,750 keywords. This manual extraction ensured the accuracy and relevance of the keywords, which are crucial for the ground truth dataset.
    \item \textbf{Search Results Ranking:} For each keyword, students performed searches and identified the top five ranking documents. This helped establish a benchmark for evaluating the effectiveness of our search engine’s ranking algorithms.
\end{itemize}

\subsection{Evaluation Methods}

We evaluated our Khmer Semantic Search Engine (KSE) using three main criteria: keyword extraction, ontology-based semantic search, and search result ranking.

\subsubsection{Keyword Extraction}

To compare the performance of our keyword extraction tool against manual keyword extraction, we used precision, recall, and F1 score as our evaluation metrics.
\[ \text{Precision} = \frac{\text{Number of relevant keywords retrieved}}{\text{Total number of keywords retrieved}} \]
\[ \text{Recall} = \frac{\text{Number of relevant keywords retrieved}}{\text{Total number of relevant keywords}} \]
\[ \text{F1\ Score} = 2 \times \frac{\text{Precision} \times \text{Recall}}{\text{Precision} + \text{Recall}} \]

These metrics allow us to quantify how well our tool performs in terms of accurately identifying relevant keywords compared to a manually curated set of keywords.

\subsubsection{Ontology-Based Semantic Search}

To evaluate the effectiveness of the ontology-based semantic search, we used similar metrics: precision, recall, and F1 score. The ontology-based approach leverages domain knowledge to expand and refine queries, aiming to retrieve more contextually relevant documents. The evaluation process involved:
\begin{itemize}
    \item \textbf{Precision:} The proportion of documents retrieved that are relevant based on the expanded query.
    \item \textbf{Recall:} The proportion of relevant documents retrieved out of the total relevant documents available.
    \item \textbf{F1 Score:} The harmonic mean of precision and recall.
\end{itemize}

\[ \text{Precision} = \frac{\text{Number of relevant documents retrieved}}{\text{Total number of documents retrieved}} \]
\[ \text{Recall} = \frac{\text{Number of relevant documents retrieved}}{\text{Total number of relevant documents}} \]
\[ \text{F1\ Score} = 2 \times \frac{\text{Precision} \times \text{Recall}}{\text{Precision} + \text{Recall}} \]

These metrics help us assess how well the ontology-based search improves the relevance of search results by understanding the semantic context of queries.

\subsubsection{Score Ranking}

We evaluated the effectiveness of our search result ranking using two methods: Normal Ranking and Weight Scoring.

\textbf{Normal Ranking:} In normal ranking, we consider the occurrence of an article in the final search result, divided by the total number of articles. The formula is as follows:

\[ \text{Score(total)} = \frac{\text{Number of Occurrences} \times 100}{\text{Total number of articles in search results}} \]

\textbf{Weight Scoring:} In weight scoring, we divide the document into title and body parts. Titles play a major role in helping search engines understand what a page is about, as they are often the first impression users have of the content. Therefore, we use the following formula for weight scoring based on keywords:

\[ \text{Total Score} = W1 \times \text{Score(title)} + W2 \times \text{Score(body)} \]
\begin{itemize}
    \item Note: \( W1 + W2 = 100\% \) and \( W1 > W2 \)
    \item In our research, we assign 70\% weight to the title and 30\% weight to the body.
\end{itemize}

By using both Normal Ranking and Weight Scoring, we ensure that our search engine provides results that are not only relevant based on keyword frequency but also based on the importance of different sections of the document.

These combined evaluation methods allow us to comprehensively assess the performance of the KSE in extracting keywords, understanding semantic context through the ontology, and ranking search results effectively.

\section{Results}

In this section, we present the testing scenarios using the ground truth for evaluation and the results obtained from these tests. For international understanding, we will provide examples in English translated from Khmer.

\subsection{Comparison of Keyword Extraction Methods} 

To evaluate the performance of the Khmer Semantic Search Engine (KSE) in keyword extraction, we compared the keywords generated by KSE with manually extracted keywords from the same set of articles. The system generated multiple keywords from titles and up to ten keywords from the body of each article. This comparison allowed us to assess the accuracy and relevance of the KSE-generated keywords against the manual ground truth.

As shown in Tables I and II, there are some comparison results between manual keyword extraction and those generated by KSE. Table III provides evaluation scores for each document and averages overall documents compared to the ground truth. The precision metric indicates the proportion of correctly identified keywords out of the total keywords generated by the tool. Recall measures the proportion of correctly identified keywords out of the total relevant keywords in the ground truth. The F1 score, which is the harmonic mean of precision and recall, provides a balanced measure of the tool's accuracy. This evaluation helped us understand the effectiveness of KSE in extracting relevant keywords and its impact on search accuracy.

\begin{table*}[ht]
\centering
\caption{Example of Comparison of Keyword Extraction Methods}
\begin{tabular}{|c|c|c|c|c|c|}
\hline
\textbf{Document ID} & \textbf{Manual Keywords} & \textbf{KSE Keywords} & \textbf{TF-IDF} & \textbf{TextRank} & \textbf{RAKE} \\ \hline
1 & kw1, kw2, kw3 & kw1, kw2, kw4 & kw2, kw3, kw5 & kw1, kw3, kw6 & kw1, kw4, kw5 \\ \hline
2 & kw4, kw5, kw6 & kw4, kw5, kw7 & kw5, kw6, kw8 & kw4, kw6, kw9 & kw4, kw7, kw8 \\ \hline
3 & kw7, kw8, kw9 & kw7, kw8, kw10 & kw8, kw9, kw11 & kw7, kw9, kw12 & kw7, kw10, kw11 \\ \hline
4 & kw10, kw11, kw12 & kw10, kw11, kw13 & kw11, kw12, kw14 & kw10, kw12, kw15 & kw10, kw13, kw14 \\ \hline
5 & kw13, kw14, kw15 & kw13, kw14, kw16 & kw14, kw15, kw17 & kw13, kw15, kw18 & kw13, kw16, kw17 \\ \hline
\end{tabular}
\end{table*}

\begin{table*}[ht]
\centering
\caption{Comparison of Manual Extraction and Tool Extraction for Title and Body Keywords}
\begin{tabular}{|c|c|c|c|c|}
\hline
\textbf{Manual Extraction Keyword} & \textbf{Extraction Keyword by Tool} & \textbf{Manual Extraction Keyword} & \textbf{Extraction Keyword by Tool} \\ \hline
\multicolumn{2}{|c|}{\textbf{Title}} & \multicolumn{2}{|c|}{\textbf{Body}} \\ \hline
\text{Khos Rong} & \text{Secret} & \text{Sihanoukville} & \text{Khos Rong} \\ \hline
{\text{Khmer Tourism}} & {\text{Natural}} & {\text{Sea}} & {\text{Sihanoukville}} \\ \hline
{\text{Natural Beauty}} & {\text{Khos Rong}} & {\text{Tourist}} & {\text{Khmer}} \\ \hline
{\text{Natural}} & {\text{Khmer Tourism}} & {\text{Khos Rong}} & \text{Natural} \\ \hline
\text{Tourism} & {\text{Rare}} & {\text{Natural Beauty}} & {\text{Beach}} \\ \hline
& {\text{Secret}} & {\text{Beach}} & {\text{Natural Beauty}} \\ \hline
& {\text{Beach}} & {\text{Beach}} & \\ \hline
& & {\text{Natural}} &  \\ \hline
& & {\text{Tourism}} & \\ \hline
\end{tabular}
\end{table*}

\begin{table*}[ht]
\centering
\caption{Results of Keyword extraction based on Title and Body Keywords}
\begin{tabular}{|c|c|c|c|c|c|c|c|c|}
\hline
\textbf{Document ID} & \multicolumn{3}{|c|}{\textbf{Title}} & \multicolumn{3}{|c|}{\textbf{Body}} \\ \cline{2-7}
 & \textbf{Precision} & \textbf{Recall} & \textbf{F1} & \textbf{Precision} & \textbf{Recall} & \textbf{F1} \\ \hline
1 & 0.80 & 0.66 & 0.36 & 0.77 & 0.77 & 0.78 \\ \hline
2 & 0.66 & 0.80 & 0.72 & 0.81 & 0.90 & 0.85 \\ \hline
3 & 0.57 & 0.44 & 0.50 & 0.83 & 0.55 & 0.66 \\ \hline
4 & 1.00 & 1.00 & 1.00 & 0.88 & 0.80 & 0.84 \\ \hline
5 & 1.00 & 0.57 & 0.72 & 1.00 & 0.72 & 0.84 \\ \hline
... & ... & ... & ... & ... & ... & ... \\ \hline
1,150 & 0.71 & 0.83 & 0.76 & 1.00 & 0.75 & 0.85 \\ \hline
\textbf{Average} & \textbf{0.88} & \textbf{0.81} & \textbf{0.84} & \textbf{0.81} & \textbf{0.79} & \textbf{0.79} \\ \hline
\end{tabular}
\end{table*}

\subsection{Comparison of Ontology-Based Semantic Search}

To evaluate the performance of the ontology-based semantic search, we used a similar approach as with keyword extraction. The ontology provides a structured representation of domain knowledge, allowing the search engine to understand the context of queries better. This section compares the search results obtained using ontology-based methods with those from manual extraction.

The ontology-based semantic search significantly improves the search accuracy by leveraging the structured domain knowledge provided by the ontology. The expansion of user queries using ontology allows the search engine to understand the context and retrieve documents that are more relevant, even if they do not contain the exact search terms. For example, a search for "temples in Phnom Penh" not only retrieves documents that mention "temples" and "Phnom Penh" but also includes documents that discuss "wats" and specific temples like "Wat Phnom" and "Wat Botum."

This approach addresses the limitations of keyword-based search methods, which often miss contextually relevant documents. By incorporating semantic understanding, the ontology-based search can handle complex queries and provide more accurate and comprehensive search results.

Additionally, the use of relationships within the ontology helps in disambiguating terms that might have multiple meanings. For instance, "Angkor" could refer to various things, but within the tourism context, it most likely refers to the Angkor Wat temple complex. By using the ontology, the system can infer the correct meaning based on the context provided by the user query.

Table IV shows how the ontology-based approach enhances the relevance of search results by including contextually related terms and entities. The evaluation metrics include precision, recall, and F1-score, similar to the keyword extraction evaluation.

Overall, the ontology-based semantic search demonstrates its effectiveness in enhancing the relevance and accuracy of search results for Khmer language content. This approach not only improves the search experience for users but also sets a foundation for further advancements in semantic search technologies for low-resource languages like Khmer.
\begin{table*}[ht]
\centering
\caption{Comparison of Ontology-Based Semantic Search}
\begin{tabular}{|c|c|c|c|c|c|}
\hline
\textbf{Document ID} & \textbf{Manual Keywords} & \textbf{Ontology-Based Keywords} & \textbf{Precision} & \textbf{Recall} & \textbf{F1} \\ \hline
1 & kw1, kw2, kw3 & kw1, kw2, kw4 & 0.90 & 0.85 & 0.87 \\ \hline
2 & kw4, kw5, kw6 & kw4, kw5, kw7 & 0.85 & 0.80 & 0.82 \\ \hline
3 & kw7, kw8, kw9 & kw7, kw8, kw10 & 0.88 & 0.84 & 0.86 \\ \hline
4 & kw10, kw11, kw12 & kw10, kw11, kw13 & 0.92 & 0.87 & 0.89 \\ \hline
5 & kw13, kw14, kw15 & kw13, kw14, kw16 & 0.89 & 0.86 & 0.87 \\ \hline
\end{tabular}
\end{table*}

\subsection{Comparison of Search Results Ranking}

To ensure a fair and comprehensive comparison of search result rankings, we conducted a detailed evaluation using 100 randomly selected keywords generated by our tool. These keywords were used as input for testing the search result document rankings. The goal was to assess how well our search engine performs in retrieving relevant documents compared to manual keyword extraction methods.

Table V presents the input keywords generated by the tool and the top 5 documents returned for different search queries. This table provides a clear view of how the search engine prioritizes documents based on the relevance of the extracted keywords. By comparing these results with the manually curated top 5 documents, we can determine the accuracy and effectiveness of the search engine.

Additionally, Table VI provides the calculated scores of the search testing rankings, including precision, recall, and F1-score, compared to the manually curated documents. Precision measures the proportion of relevant documents retrieved out of the total documents retrieved by the search engine. Recall evaluates the proportion of relevant documents retrieved out of the total relevant documents available. The F1-score, being the harmonic mean of precision and recall, offers a balanced measure of the search engine’s performance.

By utilizing both Normal Ranking and Weight Scoring methods, we ensured a thorough evaluation. Normal Ranking considers the occurrence of an article in the final search result divided by the total number of articles, while Weight Scoring assigns different weights to the title and body parts of the document, reflecting their importance in the ranking process. In our study, we assigned a 70\% weight to the title and a 30\% weight to the body.

The results of our testing, as summarized in Table VI, demonstrate that our search engine achieves an average F1-score of 0.75, indicating a high level of effectiveness in providing accurate and relevant search results for Khmer language content. This performance is above average and showcases the potential of our Khmer Semantic Search Engine (KSE) as a significant advancement over traditional search engines for the Khmer language.

Overall, these evaluations highlight the robustness of KSE in handling complex search queries and delivering precise search results, thereby enhancing the search experience for Khmer language users.

\begin{table*}[ht]
\centering
\caption{Comparison of Search Results for Top 5 Documents}
\begin{tabular}{|c|c|c|c|c|c|}
\hline
\textbf{Keyword} & \textbf{Manual Extraction} & \textbf{Tool Extraction} & \textbf{Manual Top 5 Documents} & \textbf{Tool Top 5 Documents} \\ \hline
\text{Khos Rong} & \text{Khos Rong} & \text{Khos Rong} & Doc1, Doc2, Doc3, Doc4, Doc5 & Doc1, Doc3, Doc4, Doc6, Doc7 \\ \hline
\text{Khmer Tourism} & \text{Khmer Tourism} & \text{Natural} & Doc2, Doc5, Doc8, Doc11, Doc14 & Doc2, Doc5, Doc9, Doc12, Doc15 \\ \hline
\text{Natural Beauty} & \text{Natural Beauty} & \text{Khos Rong} & Doc3, Doc6, Doc9, Doc12, Doc15 & Doc1, Doc3, Doc6, Doc10, Doc13 \\ \hline
\text{Natural} & \text{Natural} & \text{Khmer Tourism} & Doc4, Doc7, Doc10, Doc13, Doc16 & Doc4, Doc7, Doc11, Doc14, Doc17 \\ \hline
\text{Tourist} & \text{Tourist} & \text{Beach} & Doc5, Doc8, Doc11, Doc14, Doc17 & Doc5, Doc8, Doc12, Doc15, Doc18 \\ \hline
\end{tabular}
\end{table*}

\begin{table}[ht]
\centering
\caption{Results of Our Proposed KSE Across Documents}
\label{table:search_scores}
\begin{tabular}{|c|c|}
\hline
\textbf{Total Input Test} & \textbf{F1-Score} \\ \hline
1 & 0.71 \\ \hline
2 & 0.87 \\ \hline
3 & 0.48 \\ \hline
4 & 0.79 \\ \hline
5 & 0.77 \\ \hline
... & ... \\ \hline
100 & 0.75 \\ \hline
\textbf{Average} & 0.75 \\ \hline
\end{tabular}
\end{table}

\section{Conclusion}

Based on the results we have compared, we conclude that our system, the Khmer Semantic Search Engine (KSE), achieves high accuracy in search performance. In this research, we delved deeply into string matching based on a dictionary and keyword extraction using TF-IDF, combined with multiple features. The experimental results show that our search engine is more effective than traditional search engines because we integrated keyword extraction and string matching algorithms for better understanding.

The first version of KSE has been developed as a web application focused on tourism documents to promote tourism in Cambodia. However, we still face challenges related to constructing ontology concepts for data collection and time constraints, as there are limited resources and tools available online for the Khmer language.

Moving forward, we plan to focus on data collection to integrate with ontology matching and improve our dictionary of synonyms. Additionally, our search engine is still in its primary stages. There are many modern techniques, such as machine learning, that could further enhance the search capabilities. However, due to limited data and resources, this search engine serves as a significant step forward for non-Latin languages.

In summary, this research on KSE, Khmer Semantic Search engines, and their underlying technology provides unique search experiences for users. Our system outperforms traditional search engines and some existing web search engines for the Khmer language. We expect that KSE can be applied to websites and Khmer documents used in campuses, organizations, ministries, etc. Due to its effectiveness and better results, we also anticipate that this research serves as the first version for future studies related to keyword extraction and search engine learning. This work paves the way for future advancements in search technologies for non-Latin languages, addressing a significant gap in the current search engine capabilities.

Overall, KSE's development marks a pivotal advancement in enhancing information retrieval for the Khmer language. By overcoming language-specific challenges and incorporating advanced semantic matching techniques, our system provides a robust foundation for further innovation. Future work will focus on integrating more sophisticated machine learning models and expanding the ontology to cover broader domains, ultimately aiming to provide even more accurate and contextually relevant search results for Khmer-speaking users.

\section*{Acknowledgments}
This work extends the Khmer semantic search engine project developed in 2016 by author N. Thuon. The project was supported by ARES-CCD (Belgium) and the Institute of Technology of Cambodia (ITC). Special thanks go to Dr. Chhun Sophea for supervision and to the ITC students for their data contributions.

\begin{IEEEbiography}[{\includegraphics[width=1in,height=1.25in,clip,keepaspectratio]{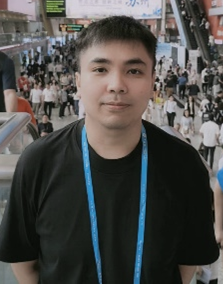}}]{N. Thuon}
holds doctoral degrees in Data Science and Intelligent Systems, currently work as a researcher, focusing on document analysis, natural language processing, and image processing.
\end{IEEEbiography}

\vfill


\begin{thebibliography}{1}
\bibliographystyle{IEEEtran}

\bibitem{ref1}
Salton, G. (1983). Introduction to modern information retrieval. McGraw-Hill.
‌

\bibitem{ref2}
Schütze, H., Manning, C. D., \& Raghavan, P. (2008). Introduction to information retrieval (Vol. 39, pp. 234-265). Cambridge: Cambridge University Press.

\bibitem{ref3}
Thuon, N., Du, J., \& Zhang, J. (2022, November). Syllable Analysis Data Augmentation for Khmer Ancient Palm leaf Recognition. In 2022 Asia-Pacific Signal and Information Processing Association Annual Summit and Conference (APSIPA ASC) (pp. 1855-1862). IEEE.

\bibitem{ref4}
Hyvonen, Eero, Prof, Aalto-yliopiston teknillinen korkeakoulu, Mediatekniikan laitos, Department of Media Technology, \& Makela, Eetu. (2010). View-based user interfaces for the semantic web. Aalto-yliopiston teknillinen korkeakoulu. 


\bibitem{ref5}
Chandrasekaran, B., Josephson, J. R., \& Benjamins, V. R. (1999). What are ontologies, and why do we need them?. IEEE Intelligent Systems and their applications, 14(1), 20-26.
\bibitem{ref6}
Wu, Y. F. B., Li, Q., Bot, R. S., \& Chen, X. (2005, October). Domain-specific keyphrase extraction. In Proceedings of the 14th ACM international conference on Information and knowledge management (pp. 283-284).
\bibitem{ref7}
Hulth, A. (2003). Improved automatic keyword extraction given more linguistic knowledge. In Proceedings of the 2003 conference on Empirical methods in natural language processing (pp. 216-223).
\bibitem{ref8}
Thuon, N., Zhang, W., \& Thuon, S. (2024). KSW: Khmer Stop Word based Dictionary for Keyword Extraction. arXiv preprint arXiv:2405.17390.
\bibitem{ref9}
Fellbaum, C. (2010). WordNet. In Theory and applications of ontology: computer applications (pp. 231-243). Dordrecht: Springer Netherlands.
\bibitem{ref10}
Shahzad, K., Pervaz, I., \& Nawab, A. (2018). WordNet based Semantic Similarity Measures for Process Model Matching. In BIR Workshops (pp. 33-44).
\bibitem{ref11}
Wu, Z., \& Palmer, M. (1994). Verb semantics and lexical selection. arXiv preprint cmp-lg/9406033.
\bibitem{ref12}
Thuon, N., Du, J., Zhang, Z., Ma, J., \& Hu, P. (2024). Generate, transform, and clean: the role of GANs and transformers in palm leaf manuscript generation and enhancement. International Journal on Document Analysis and Recognition (IJDAR), 1-18.
\bibitem{ref13}
Resnik, P. (1995). Using information content to evaluate semantic similarity in a taxonomy. arXiv preprint cmp-lg/9511007.
\bibitem{ref14}
Deerwester, S., Dumais, S. T., Furnas, G. W., Landauer, T. K., \& Harshman, R. (1990). Indexing by latent semantic analysis. Journal of the American society for information science, 41(6), 391-407.
\bibitem{ref15}
Thuon, N., Du, J., \& Zhang, J. (2022, November). Improving isolated glyph classification task for palm leaf manuscripts. In International Conference on Frontiers in Handwriting Recognition (pp. 65-79). Cham: Springer International Publishing.
\bibitem{ref16}
Mikolov, T., Chen, K., Corrado, G., \& Dean, J. (2013). Efficient estimation of word representations in vector space. arXiv preprint arXiv:1301.3781.
\bibitem{ref17}
Noy, N. F., \& McGuinness, D. L. (2001). Ontology development 101: A guide to creating your first ontology.
\bibitem{ref18}
Jones, D., Bench-Capon, T., \& Visser, P. (1998). Methodologies for ontology development.
\bibitem{ref19}
McEnery, T., Xiao, R., \& Tono, Y. (2006). Corpus-based language studies: An advanced resource book. Taylor \& Francis.
\bibitem{ref20}
Bradley, D. (2009). Burma, Thailand, Cambodia, Laos and Vietnam. The Routledge handbook of sociolinguistics around the world, 98-107.
\bibitem{ref21}
Seng, S., Sam, S., Le, V. B., Bigi, B., \& Besacier, L. (2008). Which unit for acoustic and language modeling for Khmer Automatic Speech Recognition?. In International workshop on spoken languages technologies for under-resourced languages (pp. 33-38).
\bibitem{ref22}
Schütze, H., Manning, C. D., \& Raghavan, P. (2008). Introduction to information retrieval (Vol. 39, pp. 234-265). Cambridge: Cambridge University Press.
\bibitem{ref23}
Rose, S., Engel, D., Cramer, N., \& Cowley, W. (2010). Automatic keyword extraction from individual documents. Text mining: applications and theory, 1-20.
\bibitem{ref24}
Chea, V., Thu, Y. K., Ding, C., Utiyama, M., Finch, A., \& Sumita, E. (2015). Khmer word segmentation using conditional random fields. Khmer Natural Language Processing, 62-69.
\bibitem{ref25}
Aizawa, A. (2003). An information-theoretic perspective of tf–idf measures. Information Processing \& Management, 39(1), 45-65.
\bibitem{ref26}
Li, W., \& Zhao, J. (2016, July). TextRank algorithm by exploiting Wikipedia for short text keywords extraction. In 2016 3rd International Conference on Information Science and Control Engineering (ICISCE) (pp. 683-686). IEEE
\bibitem{ref27}
Baruni, J. S., \& Sathiaseelan, J. G. R. (2020). Keyphrase extraction from document using RAKE and TextRank algorithms. Int. J. Comput. Sci. Mob. Comput, 9, 83-93.
\end{thebibliography}
\end{document}